\documentstyle[aps,prl,multicol,epsf]{revtex}

\newcommand{\bleq}{\ifpreprintsty
                   \else
                   \end{multicols}\vspace*{-3.5ex}{\tiny
                   \noindent\begin{tabular}[t]{c|}
                   \parbox{0.493\hsize}{~} \\ \hline \end{tabular}}
                   \fi}
\newcommand{\eleq}{\ifpreprintsty
                   \else
                   {\tiny\hspace*{\fill}\begin{tabular}[t]{|c}\hline
                    \parbox{0.49\hsize}{~} \\
                    \end{tabular}}\vspace*{-2.5ex}\begin{multicols}{2}
                    \fi}
\newcommand{\bcols}{\ifpreprintsty\else\begin{multicols}{2}\fi}
\newcommand{\ecols}{\ifpreprintsty\else\end{multicols}\fi}

\begin{document}
\bibliographystyle{prsty}
\title{Coulomb Drag in Systems with Tunneling  Bridges}
  
\draft

\author{Yuval Oreg$^a$, and Alex Kamenev$^b$}

\address{ $^{a}$Department of Condensed Matter Physics The Weizmann
  Institute of Science 76100 Rehovot, Israel \\
  $^{b}$Institute for Theoretical Physics University of California
  Santa Barbara, CA 93106-4030, U.S.A.
  \\
  {}~{\rm (\today)}~
  \medskip \\
  \parbox{14cm} {\rm We study the Coulomb drag effect in double layer
    electronic systems with local tunneling links. The possibility of
    tunneling between the layers leads to a pronounced exchange
    contribution to the transconductance, which is negative and
    non--vanishing at zero temperature.  The diffusive motion of the
    electrons inside each layer in interplay with the
    electron--electron interaction cause a distinguishable singular
    temperature dependence of the transconductance at low
    temperatures.
    \smallskip\\
    PADS numbers: 73.20.Dx, 73.40.Gk, 73.50.Yg }\bigskip \\ }

\maketitle

\bcols Coulomb drag in spatially separated double--layer electronic
systems has received a great deal of attention recently, both
experimentally
\cite{DR:Solomon89,DR:Gramila91,DR:Sivan92,DR:Gramila93,DR:Gramila94}
and theoretically
\cite{DR:Laikhtman90,DR:Jauho93,DR:Zheng93,DR:Kamenev95,DR:Flensberg95,DR:Ussishkin97,DR:Kim96}.
A setup for studying the drag effect consists of applying current to
one of the two layers, and measuring the induced voltage in the second
layer. The ratio between the two, known also as the {\em
  transresistance}, $\rho_{D}$, is the main characteristic of the drag
effect. Theoretically, it is more convenient to study the {\em
  transconductance}, $\sigma _{D}$, which is related to
transresistance as (for $\sigma _{D}\ll \sigma $)
\begin{equation}
\rho_{D}= - \sigma_{D}/ \sigma^2 ,  \label{eq:rhosigma}
\end{equation}
where $\sigma = e^2 D\nu $ is the conductance of a single isolated
layer (hereafter we discuss the case of identical layers); $D$ is the
diffusion constant, and $\nu $ is the density of states per unit area.
Most of the previous studies were concentrated on systems, in which
the coupling between the layers is only due to the interlayer Coulomb
interaction.  In this case density fluctuations of the electron liquid
in the first layer induce fluctuations in the second one, which in
turn lead to an induced current
\cite{DR:Laikhtman90,DR:Jauho93,DR:Zheng93,DR:Kamenev95,DR:Flensberg95}.
Since it involves classical thermal fluctuations, interacting in a
Hartree way, the drag effect vanishes at zero temperature. Moreover,
the sign of the effect does not depend on the nature of the
interaction (repulsion or attraction) and has to do only with the sign
of the charge carriers in the two layers.  For example, for carriers
of the same type, the drag transconductance is strictly {\em positive},
i.e., the induced current flows in the direction of the applied one.

In this paper we address a qualitatively different mechanism which
leads to a transconductance. It takes place in the presence of
pointlike shortages (or bridges) --- points where electrons may tunnel
between the two layers.  Such bridges are often present in metallic
double--layer systems.  As we show below the interplay between
tunneling and the Coulomb interactions lead to a transconductance
which is non-vanishing at zero temperature and {\em negative} (for
carriers of the same charge). That is, the induced current flows in
the direction opposite to the driving one.  The origin of the effect
is in the {\em exchange} interaction between the electrons of
different layers, which is possible due to the wavefunctions' overlap
at the bridges. To illustrate the physics of the effect consider a
wavepacket propagating in the first layer in the direction of an
applied current.  Once it reaches a bridge, a part of the wavepacket
tunnels into the second layer, while the remaining part continues to
move in the initial direction. Without interlayer electron--electron
($e$--$e$) interaction, the wavefront propagating in the second layer
is spherically symmetric, thus no net current is induced.  (Indeed,
after tunneling through a point contact the electron completely
``forgets'' the initial direction of its momentum.)  In the presence
of interaction the wavefront moves in a direction that minimized its
interaction with the initial wavepacket.  Thus, for a repulsive
interaction the electrons that tunneled move in a direction opposite
to that before tunneling. This leads to a {\em negative}
transconductance.  As we shall see below such a lowest order (in
tunneling) mechanism is dominant only at not too small temperatures.
At lower temperatures the leading mechanism involves coherent
tunneling to the second layer and back to the first one accompanied by
intralayer Coulomb interactions.  The exchange contribution, unlike
the Hartree term, is not proportional to a small electron--hole
asymmetry factor
\cite{DR:Laikhtman90,DR:Jauho93,DR:Zheng93,DR:Kamenev95,DR:Flensberg95},
therefore, the former's absolute value may well overcome the latter's even
for a small tunneling rate.  In the case of diffusively propagating
wavepackets the exchange transconductance has a peculiar temperature
dependence, which is described below.
 
The tunneling rate between the two layers may be described by a mean
intralayer lifetime, $\tau_{12}$, which is related to the {\em
  interlayer} tunneling conductance per unit area, $\sigma_{\perp}$,
as
\begin{equation}
\label{eq:tau12} 
\sigma_{\perp} = e^2  \nu/\tau_{12} \, .
\end{equation} 
We shall assume that the tunneling is weak, i.e., $\tau_{12}\gg \tau$,
where $\tau$ is the mean elastic scattering time within each layer.
Three energy scales determine the temperature dependence of the
transconductance: $\hbar/\tau_{12} < (\kappa
d)\hbar/\tau_{12} < \hbar/\tau$ \cite{DR:foot1}, where $\kappa=2\pi
e^2\nu$ is the inverse Thomas--Fermi screening radius, and $d$ is the
distance between the layers.  

At high temperatures, $\hbar/\tau < T$, the motion of the wavepackets
is ballistic and the exchange contribution to the transconductance is
negative and temperature independent [its value is discussed below,
see Eq.~(\ref{eq:ball})].  At $T<\hbar/\tau $ the diffusive character
of the electron motion should be taken into account. Up to the lowest
order in the tunneling amplitude we find the following temperature
dependent contribution to the transconductance
\begin{mathletters}
\label{sigma_D}
\begin{equation}
\sigma_{D}= - \frac{e^2}{\hbar} \frac{1}{24 \pi}\frac{\ln (\kappa d) }{\kappa d}
\frac{1}{T\tau_{12}}.
                                                                     \label{eq:dirty}
\end{equation}
The appearance of a contribution which has a singular temperature
dependence is not entirely unexpected for a diffusive system
\cite{DS:Altshuler79}.  What is less common is the fact that the
divergence is so pronounced, $\sim T^{-1}$, instead of being
logarithmic, as in the case of Altshuler--Aronov interaction
corrections to the conductivity of 2D systems \cite{DS:Altshuler79}.
The temperature dependence becomes even more singular at smaller
temperatures, $T < (\kappa d)\hbar/\tau_{12}$ \cite{DR:foot1}
\begin{equation}
\sigma_{D}=- \frac{e^2}{\hbar} \frac{3 \zeta (3)}{8 \pi^4}
\frac{\ln(T \tau_{12})}{(T\tau_{12})^2}\, .
                                                       \label{eq:intra}
\end{equation}
The temperature divergences in Eqs. (\ref{eq:dirty}) and (\ref{eq:intra})
are cut off by the system size, $L$, at the Thouless temperature,
$\hbar D/L^2$. For systems larger than the mean propagation distance
within one layer, $\sqrt{D \tau_{12}}$, there is an additional
temperature range $\hbar D/L^2 < T < \hbar/\tau_{12}$. Here the
exchange contribution to the transconductance is dominated by multiple
tunneling processes, leading to the form
\begin{equation}
\sigma _{D} = -\frac{e^2}{\hbar} \frac{1}{8\pi^2}  
\ln \left(T \tau_{12} \right)^{-1}\,  .  
                                                               \label{eq:mult}
\end{equation}
\end{mathletters}
This expression differs from the Altshuler--Aronov interaction
correction to the 2D conductivity \cite{DS:Altshuler79} only by a
factor of $1/4$, which reflects the peculiarity of the drag
measurement setup. Where the current flows only in one half of the
system and the potential is measured in the other half.

We now turn to the derivation of
Eqs.~(\ref{eq:dirty})--(\ref{eq:mult}).  The Hamiltonian of the system
is given by
\begin{equation}
H=H_{1}+H_{2}+H_{\mbox{\footnotesize int}}+H_{T}\,,  \label{H}
\end{equation}
where $H_{1(2)}$ is the Hamiltonian of the first (second) isolated
layer, including elastic disorder, and $H_{\mbox{\footnotesize int}}$
includes interlayer as well as intralayer Coulomb interactions. The
first three terms on the r.h.s. of Eq.~(\ref{H}) are traditionally
involved in the description of the drag effect 
\cite{DR:Zheng93,DR:Kamenev95,DR:Flensberg95}. We add the term describing
pointlike tunneling processes
\begin{equation}
H_{T} = V \sum\limits_{i=1}^{N} \sum\limits_{k,p} 
e^{i r_i(k-p)} a_{k}^{\dagger}b_{p} +  \mbox{h.c.}  \, ,  
                                                               \label{HT}
\end{equation}
where $a(a^\dagger)$ and $b(b^\dagger)$ are the annihilation
(creation) operators of electrons in the first and the second layers
respectively, and $r_i$, $i=1\ldots N$, are the random positions of
$N$ bridges.  The tunneling amplitude $V$ is related to the lifetime
$\tau_{12}$ as $2 \pi \nu L^2 N |V|^2 = \hbar/\tau_{12}$ (cf.
Eq.~(\ref{eq:tau12})).

\begin{figure}
\vglue 0cm
\hspace{0.01\hsize}
\epsfxsize=0.9\hsize
\epsffile{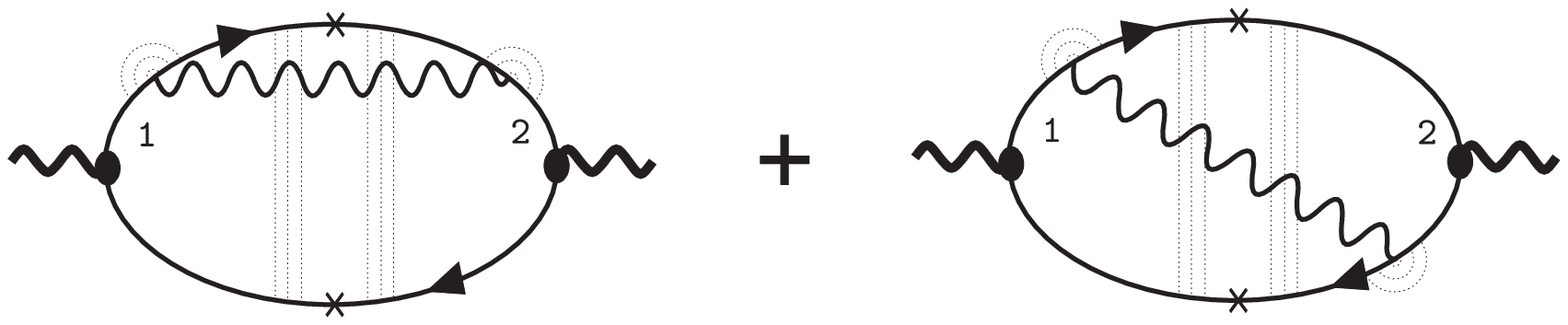}
\refstepcounter{figure} \label{fg:pinholedrag}
{\small FIG.\ \ref{fg:pinholedrag} Two diagrams contributing to the
  transconductance that are second order in tunneling (denoted by
  $\times$).  Full lines with arrows are electron Green functions,
  dashed lines represent diffusons, and wavy lines screened
  interactions.  Two additional diagrams with arrows in the opposite
  direction should be included. The numbers indicate the layer
  index.\par}
\end{figure}

At various  temperatures the exchange contribution to the 
transconductance originates from the different processes. For example, for
$(\kappa d) \hbar/\tau_{12} <T < \hbar /\tau$ the dominant
contribution is second order in tunneling and is given by
the four diffuson diagrams depicted in Fig.~\ref{fg:pinholedrag}.
Performing an analytical continuation and taking the dc limit in a
standard manner \cite{DS:Altshuler82}, one finds for the exchange contribution
to the transconductance
\begin{equation}
\sigma _{D}=
i \frac{\sigma}{4\pi} \int\limits_{-\infty }^{\infty}\!\!\!\!d\omega 
\frac{\partial }{\partial \omega }
\left[ \omega \coth \frac{\omega }{2T} \right]
F_{xx}(\omega )\,.  
                                                            \label{AA1}
\end{equation}
A straightforward summation over the fast electronic momenta  leads 
to $F_{xx}(\omega)=F_{xx}^{(\mbox{\footnotesize{a}})}(\omega)$ with
\begin{equation}
F_{xx}^{(\mbox{\footnotesize{a}})}(\omega)= \frac{8}{\tau_{12}}
\sum_{Q}DQ_{x}^2 \frac{U_{12}(Q,\omega)} {(DQ^{2}-i\omega)^{4}} \, ,  
                                                             \label{FD}
\end{equation}
where $U_{12}(Q,\omega)$ is the dynamically screened interlayer
Coulomb interaction. In the diffusive limit ($\omega< \hbar /\tau $)
it was found in Refs.~\cite{DR:Zheng93,DR:Kamenev95} (see also
Eqs.~(\ref{eq:RPA}) and (\ref{eq:pi}) below).  Substituting
Eq.~(\ref{FD}) into Eq.~(\ref{AA1}) and carefully performing
frequency and slow momenta integrations, one recovers
Eq.~(\ref{eq:dirty}).

\begin{figure}
\vglue 0cm
\hspace{0.01\hsize}
\epsfxsize=0.9\hsize
\epsffile{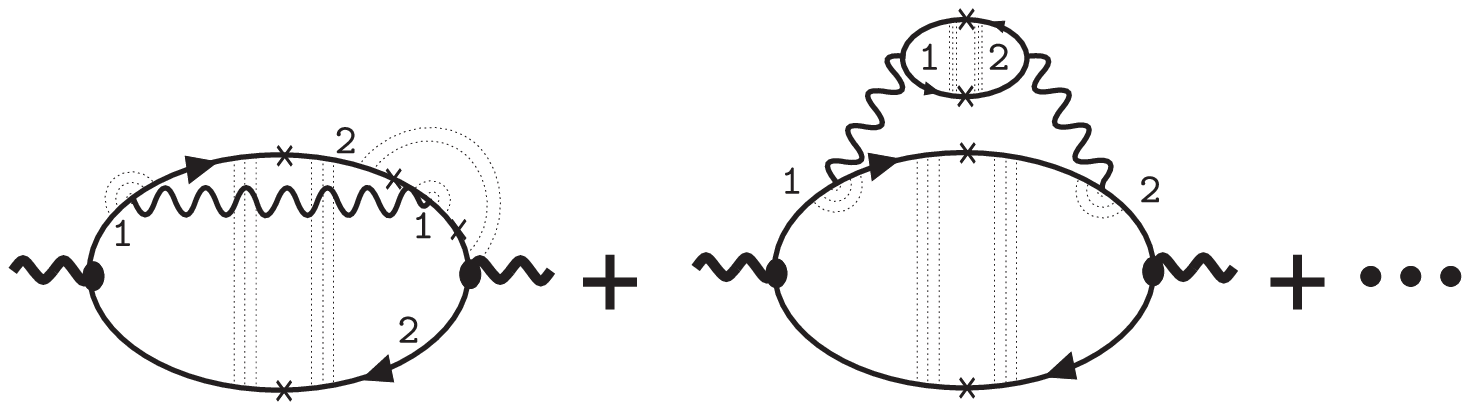}
\refstepcounter{figure} \label{fg:pinintra} {\small FIG.\ 
  \ref{fg:pinintra} Examples of diagrams contributing to the
  transconductance that are fourth order in tunneling.  Diagrams with
  interaction lines connecting ``upper'' and ``lower'' Green
  functions, as well as diagrams with an opposite direction of
  electron lines should also be included.  The numbers indicate the layer
  index.  \par}
\end{figure}

At somewhat smaller temperatures the fourth order (in the tunneling
amplitude) processes may become dominant.  Although being smaller by
the additional factor of $|V|^2\propto 1/\tau_{12}$, it contains
intra-- (instead of inter--) layer interaction, which is stronger by a
$\kappa d>1$ factor. Moreover, the fourth order term has a more
singular temperature dependence (because additional diffusons poles
are involved) which makes it eventually dominant at $T<(\kappa
d)\hbar/\tau_{12}$.  Examples of the diagrams, contributing to the
fourth order are depicted in Fig.~\ref{fg:pinintra}. The calculation
of these diagrams follows the steps outlined above and leads to the
result given in Eq.~(\ref{eq:intra}). Instead of reproducing these
calculations we develop a general formalism suitable for
multi-tunneling processes, and obtain the fourth order contribution as
a particular case.

The treatment of multiple tunneling processes through randomly placed
bridges is similar to the theory of metals with spin--flipping
impurities \cite{SOC:Hikami80}.  In our case the role of the spin $z$
component is played by the layer index.  The main difference is that
the different ``spin'' components are subjected to different (and
uncorrelated) impurity potentials.  We introduce the matrix notations
for the diffusons $D_{mn}^{kl}(Q,\omega)$, where the indices
$k,l,m,n=1,2$ indicate the layer index for each of the four incoming
and outgoing legs. Such a matrix diffuson is a solution of the following
Bette--Salpeter equation
\begin{equation}
D_{mn}^{kl} = W_{mn}^{kl} + 
\sum\limits_{i,j=1}^2 W_{mj}^{ki} \, \zeta (Q,\omega)D_{jn}^{il}\, ,  
                                                              \label{BS}
\end{equation}
where 
$\zeta (Q,\omega)=\tau^{-1} + \tau_{12}^{-1} - (DQ^{2} - i\omega) $, 
and the combined  tunneling  and impurity scattering amplitude is 
given by 
\begin{equation}
W_{mn}^{kl}=\frac{1}{\tau}\delta _{kl}\delta _{mn}\delta _{km} +
\frac{1}{\tau _{12}}(1-\delta _{kl})(1-\delta _{mn})\, .  
                                                              \label{W}
\end{equation}
Since the elastic disorder in the two layers is uncorrelated one may look 
for a solution of Eq.~(\ref{BS}) in the form
\begin{equation}
D_{mn}^{kl}(Q,\omega)=A\delta _{kl}\delta _{mn}\delta _{km}+B(1-\delta
_{kl})(1-\delta _{mn})\, ,   
                                                              \label{BS1}
\end{equation}
then in the limit $\tau_{12} \gg \tau$ one finds
\begin{equation}
A,B= \frac{1}{2} \left[ 
\frac{1}{DQ^{2}-i\omega} \pm \frac{1}{DQ^{2} -i\omega +2/\tau_{12}} \right]\, .
                                                                \label{AB}
\end{equation} 
The function $F_{xx}(\omega)$ which enters in expression (\ref{AA1}) for 
the transconductance may be written as 
(cf. the diagrams in Fig.~\ref{fg:pinholedrag}) 
\begin{equation}
F_{xx}(\omega )=4\sum\limits_{Q} 
D Q_x^2 \sum\limits_{knlm}U_{kn} 
D_{lk}^{1k}\left[ D_{12}^{lm} + D_{1m}^{l2} \right] D_{2n}^{mn}\, .  
                                                            \label{Fmult}
\end{equation}
Here $U_{kn}(Q,\omega)$ is the dynamically screened interaction 
between layers $k$ and $n$. In the RPA approximation it is 
given by a solution of 
\begin{equation}
U_{kn} = U^0_{kn} - \sum\limits_{l,m=1}^2  U^0_{kl}\Pi_{lm}U_{mn}\, , 
                                                     \label{eq:RPA}
\end{equation}
where $U^0_{kn}$ is the matrix of bare interactions
\cite{DR:Laikhtman90,DR:Jauho93,DR:Zheng93,DR:Kamenev95,DR:Flensberg95} 
and $\Pi_{kn}(Q,\omega)$ is the polarization matrix which in the present case 
takes the form 
\begin{equation}
\Pi_{kn}(Q,\omega) = \nu  \left[ 
\delta_{kn}+i\omega D^{kn}_{kn}(Q,\omega) \right]. \, 
                                                 \label{eq:pi}
\end{equation}
Note, that by virtue of Eqs.~(\ref{BS1}) and (\ref{AB}) $\Pi_{kn}$ 
satisfy the particle conservation law: $\sum_n \Pi_{kn}(Q=0,\omega)=0$.

Now, we are in the position to calculate $F_{xx}(\omega)$ for various
values of $\omega$. The second order contribution to the exchange
transconductance (Fig.~\ref{fg:pinholedrag}) comes from the expression
(cf.  Eq.~(\ref{Fmult})) $U_{12}D_{11}^{11} D_{12}^{12}
D_{22}^{22}=U_{12}A^2B$. Expanding now up to the lowest order in
$\hbar/\tau_{12}$, one obtains Eq.~(\ref{FD}).  The fourth order
originates from terms like $U_{11}D_{11}^{11} D_{12}^{12}
D_{21}^{21}= U_{11}AB^2$, which should be taken in the lowest
non-vanishing order in $\hbar/\tau_{12}$, and from the term
$U_{12}D_{11}^{11} D_{12}^{12} D_{22}^{22}$ where
$U_{12}=-U_{11}\Pi_{12}U_{22}$. As a result one obtains
$F_{xx}(\omega)=F^{(\mbox{\footnotesize{b}})}_{xx}(\omega)$ with
\begin{equation}
F^{(\mbox{\footnotesize{b}})}_{xx}(\omega)=\frac{8}{\tau_{12}^2}
\sum_Q DQ_x^2 \left[
    \frac{U_{11}+U_{22} }{(DQ^2-i\omega)^5} - 
    \frac{U_{11}\Pi_{12}U_{22}\tau_{12}}{(DQ^2-i\omega)^4}
                              \right].
                                                    \label{FDIntra1}
\end{equation}
The two terms on the r.h.s. of Eq.~(\ref{FDIntra1}) are represented by the 
two classes of diagrams exemplified in Fig.~\ref{fg:pinintra}.  
Substitution of Eq.~(\ref{FDIntra1}) in
Eq.~(\ref{AA1}) and integration  over frequency and slow momenta yields
Eq.~(\ref{eq:intra}). It is easy to see that this contribution 
is the dominant one at $T < (\kappa d)\hbar/\tau_{12}$.

Finally in the limit of $\omega <\hbar/\tau_{12}$ one may neglect the
second term in the expression for $A$ and $B$  [Eq.~(\ref{AB})] and  then
$D_{11}^{11}=D_{22}^{22}=D_{12}^{12}=D_{21}^{21}=\left( 1/2\right)
\left(DQ^{2}- i\omega \right)^{-1}$.  This yields for $\omega < 1/\tau_{12}$,
$F_{xx}(\omega)=F^{(\mbox{\footnotesize{c}})}_{xx}(\omega)$ with
\begin{equation}
F^{(\mbox{\footnotesize{c}})}_{xx}(\omega)=\sum_{Q} DQ_x^2  
\frac{U_{11} + 2U_{12}+U_{22} }{(DQ^{2}- i\omega)^{3}}\, .   
                                                       \label{Fmult1}
\end{equation}
Eqs.~(\ref{eq:RPA}) and (\ref{eq:pi}) in the present case lead to
$U_{11} + U_{12} =U_{12} + U_{22}= \nu^{-1} (DQ^{2}-i\omega)/DQ^{2}$.
As a result of slow momenta summation one obtains for $\omega
<\hbar/\tau_{12}$: $F^{(\mbox{\footnotesize{c}})}_{xx}(\omega) =
ie^2/(4\pi\sigma \omega)$.  After substitution of
$F^{(\mbox{\footnotesize{c}})}_{xx}(\omega)$ into Eq.~(\ref{AA1}) and
integrating over the frequency one obtains a logarithm
[Eq.~(\ref{eq:mult})], similar to the logarithm in
Ref.~\cite{DS:Altshuler79}.  For finite size systems the slow momentum
integration should be replaced by a discrete summation. This results
in placing the low--frequency cutoff, $\omega \sim \hbar D/L^2$, in
Eqs.~(\ref{FD}), (\ref{FDIntra1}) and (\ref{Fmult1}). As a consequence
the temperature dependence of Eqs.  ~(\ref{eq:dirty})--(\ref{eq:mult})
is flattened out at $T < \hbar D/L^2$. In order to find the exact
exchange contribution to transconductance in the transitions between
the various regimes (including the temperature independent parts) one
may directly substitute Eq.~(\ref{Fmult}) into Eq.~(\ref{AA1}) and
evaluate the resulting double integral.

Finally we briefly discuss the ballistic limit, $\hbar/\tau < T$.
For such a temperature range one should consider the diagrams depicted in
Fig. \ref{fg:pinholedrag} without the diffusons.  In this case the
function $F_{xx}=F_{xx}^{({\mbox{\footnotesize{bal}}})}$ is given by
\begin{equation}
F_{xx}^{{\mbox{\footnotesize{bal}}}}(\omega ) = \frac{4\tau^2}{D\tau_{12}} 
\sum_{Q} U_{12}(Q) \left[ \frac{Q_{x}}{Q^{2}}
\left( 1-\frac{1}{\sqrt{1+X^{2}}}\right) \right]^{2}\, ,  
                                                            \label{FB}
\end{equation}
where $X\equiv v_{F}Q/(-i\omega +1/\tau )$, and $U_{12}(Q)$ is the
screened interlayer interaction in the clean limit
\cite{DR:Laikhtman90,DR:Jauho93,DR:Zheng93,DR:Kamenev95,DR:Flensberg95}.
The slow momenta summation and frequency integration according to
Eq.~(\ref{AA1}) lead to a negative and temperature independent
transconductance, which is given  by:
\begin{equation}
\sigma_{D} =
- \frac{e^2}{\hbar} \frac{\pi}{32} \frac{1}{\kappa d} \frac{v_F \tau^2}{d \tau_{12}}\, . 
                                                                  \label{eq:ball}
\end{equation}
One should not overestimate the significance of this negative constant
exchange contribution because of the existence of two additional
``parasitic'' effects. Both contribute  positive temperature
independent constants to the measured transconductance.

The first effect is related to a possibility of momentum conserving
tunneling.  In this case the transconductance is nonzero (and
positive) at zero temperature even {\em without} any $e$--$e$
interactions.  It originates from the delocalization of the
wavefunctions between the two layers, and may be described in the
framework of the single--particle picture
\cite{DR:Boebinger91,DR:Berk94a}.  This type of (tunneling)
transconductance may be differentiated from other mechanisms by a
strong sensitivity to an {\em in-plane} magnetic field or a gate
voltage. A Fermi--surface mismatch introduced by these factors leads
to a rapid suppression of the transconductance
\cite{DR:Boebinger91,DR:Berk94a}.  The second effect originates from
the classical (Hartree) interactions between electrons. In the
presence of many bridges it may be visualized by a network of
classical resistors.  A simple calculation based on Kirchoff's laws
leads to a positive transconductance. For $L >\sqrt{D \tau_{12}}$ the
network model predicts that approximately half of the current applied
to the first layer eventually leaks out to the second one. In this
case the transresistance practically reduces to the resistance of a
single layer of a doubled width [thus it is not surprising that we
have recovered the Altshuler--Aronov interaction correction at low
temperatures and at a large system size, Eq.~(\ref{eq:mult})].
However, we stress that both mechanisms mentioned above lead to
a temperature independent effect, which is easily distinguishable from
the strong temperature dependent exchange contribution to the
transconductance.

To conclude we have examined the influence of point-like tunneling
between the two layers, in combination with $e$-$e$ interactions, on
the drag effect and the transconductance.  We found a distinguishable
exchange contribution to the transconductance, which is negative, has
a strong temperature dependence and is non-vanishing at zero
temperature.  The magnitude of the effect under optimal conditions may
reach a fraction of $e^2/2\pi\hbar$ (for the transconductance).

Discussions with A.~M.~Finkel'stein, D.~E.~Khmelnitskii,
A.~MacDonald, A. Stern and I. Ussishkin are highly acknowledged. This
research was supported by the German-Israel Foundation (GIF) and the
U.S.-Israel Binational Science Foundation (BSF), A.K. was supported by
the Rothschild fellowship.


\begin{thebibliography}{10}

\bibitem{DR:Solomon89}
P.~M. Solomon, P.~J. Price, D.~J. Frank, and D.~C.~L. Tulipe, Phys. Rev. Lett.
  {\bf 63},  2508  (1989).

\bibitem{DR:Gramila91}
T.~J. Gramila, J.~P. Eisenstein, A.~H. MacDonald, L.~N. Pfeifer, and K.~W.
  West, Phys. Rev. Lett. {\bf 66},  1216  (1991).

\bibitem{DR:Sivan92}
U. Sivan, P. M.Solomon, and H. Shtrikman, Phys. Rev. Lett. {\bf 68},  1196
  (1992).

\bibitem{DR:Gramila93}
T.~J. Gramila, J.~P. Eisenstein, A.~H. MacDonald, L.~N. Pfeifer, and K.~W.
  West, Phys. Rev. B {\bf 47},  12957  (1993).

\bibitem{DR:Gramila94}
T.~J. Gramila, J.~P. Eisenstein, A.~H. MacDonald, L.~N. Pfeifer, and K.~W.
  West, Physica B {\bf 197},  442  (1994).

\bibitem{DR:Laikhtman90}
B. Laikhtman and P.~M. Solomon, Phys. Rev. B {\bf 41},  9921  (1990).

\bibitem{DR:Jauho93}
A.-P. Jauho and H. Smith, Phys. Rev. B {\bf 47},  4420  (1993).

\bibitem{DR:Zheng93}
L. Zheng and A.~H. MacDonald, Phys. Rev. B {\bf 48},  8203  (1993).

\bibitem{DR:Kamenev95}
A. Kamenev and Y. Oreg, Phys. Rev. B {\bf 52},  7516  (1995).

\bibitem{DR:Flensberg95}
K. Flensberg, B.~Y.-K. Hu, A.-P. Jauho, and J. Kinaret, Phys. Rev. B {\bf 52},
  14761  (1995).

\bibitem{DR:Ussishkin97}
I. Ussishkin and A. Stern, Phys. Rev. B {\bf 56},  4013  (1996).

\bibitem{DR:Kim96}
Y.~-B.~Kim and A.~J.~Millis cond-mat/ 9611125.

\bibitem{DR:foot1}
Hereafter we consider a case $\kappa d \gg 1$. In a situation when $\kappa d
  \approx 1$ the regime described by Eq.~(\protect{\ref{eq:intra}}) does not
  exist. In the other extreme $(\kappa d) /\tau_{12} \gg 1/\tau$ the region of
  applicability of Eq.~(\protect{\ref{eq:dirty}}) vanishes.

\bibitem{DS:Altshuler79}
B.~L. Altshuler and A.~G. Aronov, Sol. Stat. Com. {\bf 30},  115  (1979).

\bibitem{DS:Altshuler82}
B.~L. Altshuler, A.~G. Aronov, D.~E. Khmelnitskii, and A.~I. Larkin,  in {\em
  Quantum theory of solids}, edited by I.~M. Lifshits (MIR Publishers, Moscow,
  1982).

\bibitem{SOC:Hikami80}
S. Hikami, A.~I. Larkin, and Y. Nagaoka, Prog. Theor. Phys. {\bf 63},  707
  (1980).

\bibitem{DR:Boebinger91}
G. Boebinger, A. Passner, L.~N. Pfeiffer, and K.~W. West, Phys. Rev. B {\bf
  43},  12675  (1991).

\bibitem{DR:Berk94a}
Y. Berk, A. Kamenev, a.~Palevski, L.~N. Pfeiffer, and K.~W. West, Phys. Rev. B
  {\bf 51},  2604  (1995).

\end{thebibliography}

\ecols
\end{document}